# Ferroelectric quantum Hall phase revealed by visualizing Landau level wave function interference

Mallika T. Randeria[*,1], Benjamin E. Feldman[*,1,†], Fengcheng Wu[*,2], Hao Ding[1], Andras Gyenis[1], Huiwen Ji[3], R. J. Cava[3], Allan H. MacDonald[4], Ali Yazdani[1,‡]

[1]*Joseph Henry Laboratories & Department of Physics, Princeton University, Princeton, NJ 08544, USA*
[2]*Materials Science Division, Argonne National Laboratory, Argonne, IL 60439, USA*
[3]*Department of Chemistry, Princeton University, Princeton, NJ 08544, USA*
[4]*Department of Physics, The University of Texas at Austin, Austin, TX 78712, USA*

[†]*Present address: Department of Physics and Geballe Laboratory for Advanced Materials, Stanford University, Stanford, CA 94305, USA*

*These authors contributed equally to the manuscript.
‡ email: yazdani@princeton.edu

**Novel broken symmetry states can spontaneously form due to Coulomb interactions in electronic systems with multiple internal degrees of freedom. Multi-valley materials offer an especially rich setting for the emergence of such states, which have potential electronic and optical applications[1-4]. To date, identification of these broken symmetry phases has mostly relied on the examination of macroscopic transport or optical properties. Here we demonstrate a powerful direct approach by visualizing the wave functions of bismuth surface states with a scanning tunneling microscope. Strong spin-orbit coupling on the surface of bismuth leads to six degenerate teardrop-shaped hole pockets[5]. Our spectroscopic measurements reveal that this degeneracy is fully lifted at high magnetic field as a result of exchange interactions, and we are able to determine the nature of valley ordering by imaging the broken symmetry Landau level (LL) wave functions. The remarkable spatial features of singly degenerate LL wave functions near isolated defects contain unique signatures of interference between spin-textured valleys, which identify the electronic ground state as a quantum Hall ferroelectric. Our observations confirm the recent prediction[6] that interactions in strongly anisotropic valley systems favor the occupation of a single valley, giving rise to emergent ferroelectricity in the surface state of bismuth.**

Two-dimensional (2D) systems provide an attractive platform to explore broken symmetry phases because their electronic states often possess internal degrees of freedom that are sensitive to external fields. Of particular interest is the role of valley and spin degeneracies in the quantum Hall regime, where the formation of Landau levels (LLs) allows interactions to dominate. Coulomb interactions can result in ordered phases that spontaneously break one or more symmetries of the Hamiltonian when an integer subset of the degenerate LLs is occupied.

This phenomenon, known as quantum Hall ferromagnetism[7], produces a variety of broken symmetry states across several 2D systems, including spin- and valley-polarized ground states[8-10]. Recent measurements have focused on multi-valley systems that exhibit more exotic electronic behavior. In monolayer[11] and bilayer graphene[12] as well as transition metal dichalcogenides[2], states with coupled spin and valley order have been identified based on their response to applied electric and magnetic fields. Moreover, valley polarization has been shown to lead to novel behavior such as nematic electronic order in systems with anisotropic valleys[13-15]. A recent proposal suggests that valley polarization can also result in a new type of broken symmetry quantum Hall phase: a ferroelectric ground state in which the Landau orbits have an intrinsic in-plane dipole moment[6]. These states can arise in a number of materials[5,16] which contain individual valleys that lack two-fold rotational symmetry, resulting in LL wave functions that are not inversion symmetric. Although conventional ferroelectricity has been widely studied in thin films[17-19], to date there have been no experimental reports of a quantum Hall ferroelectric phase in any material system.

In this work, we study the Bi(111) surface, which offers an especially rich arena to explore various valley ordered electronic states. Strong spin-orbit coupling splits two surface state bands to produce multiple electron and hole Fermi pockets (Fig. 1a)[5,20-23]. We focus on the six degenerate teardrop-shaped hole valleys, each of which has a distinct nontrivial spin texture[24-26] (examples for two valleys are shown in Fig. 1b-c). Due to spin-orbit splitting, there is no remaining spin degeneracy, so the only tunable quantum degree of freedom is the valley index. We define an effective filling factor $\tilde{v}$, which ranges from zero to six, as the number of occupied hole LLs within a given orbital index. Previous STM measurements have probed the valley ordering of these hole states in the quantum Hall regime, and showed that a combination of strain and exchange interactions result in doubly degenerate LLs arising from subsets of valleys with the same orientation[15]. These states break the rotational symmetry of the underlying crystal lattice, giving rise to nematic order at even filling factors. At odd integer filling, the Bi(111) surface states have the potential to exhibit a further symmetry breaking by lifting the remaining degeneracy between pairs of valleys at opposite momenta to produce a new class of quantum Hall state. There are two possibilities for such a ground state at $\tilde{\nu} = 1$[6, 27]: a coherent superposition of different valleys, which gives rise to periodic charge density modulations or a valley-polarized state that is ferroelectric because each hole valley lacks twofold rotational symmetry around its center. Recent theoretical work shows that the long-range part of the Coulomb interaction favors occupation of a single valley[6], but this valley ordering has yet to be probed experimentally.

To address the nature of the electronic ground state, we use a dilution refrigerator STM[28] to perform LL spectroscopy of the Bi(111) surface, which allows us to resolve broken symmetry quantum Hall states at all integer filling factors. High resolution wave function mapping of spontaneously formed singly degenerate LLs reveals intricate interference patterns that reflect mixing between individual electronic states in different valleys. By developing a comprehensive theoretical model and comparing it to our data, we determine that the ground state at odd filling factors is a valley-polarized quantum Hall ferroelectric phase.

We first focus on spectroscopic measurements, which demonstrate a complete lifting of the valley degeneracy of the Bi(111) surface state LLs as the filling factor is tuned by changing magnetic field. The evolution of the differential conductance $G$ as a function of magnetic field $B$ in Fig. 1d reveals an intricate pattern of energy gaps arising from a combination of single-particle and many-body effects. We focus on the dispersion of the hole-like LLs with orbital index $N = 3$ and $N = 4$, and observe that the (6)-fold valley degeneracy of these states is lifted, producing (2)- and (4)-fold degenerate LLs; the splitting between these multiplets is present at all magnetic fields, and is caused by local strain[15]. Within each multiplet, Coulomb interactions further split the LLs as they cross the Fermi level. For instance, between 12.2 and 13.3 T, several additional energy gaps open and close within the (4)-fold degenerate multiplet of the $N = 3$ LL when it is pinned to the Fermi level, also shown in the corresponding spectra at particular magnetic fields in Fig. 1e. Two exchange-split LL peaks are visible throughout this entire field range, and their relative amplitudes and energies change as a function of magnetic field. This behavior indicates the formation of multiple distinct broken-symmetry phases as the number of occupied hole-like LLs within a given orbital index takes on all integer values from $\tilde{v} = 2$ to $\tilde{v} = 6$. Similar behavior is also visible when the (2)-fold degenerate multiplet of the $N = 4$ LL splits into two singly degenerate states (at $\tilde{v} = 1$) as it crosses the Fermi level around $B = 11.1$ T (Fig. 1f). These states at odd integer filling factors go beyond previous spectroscopic measurements, which showed only a partial lifting of the valley degeneracy into three doubly degenerate LLs[15].

Splitting within each multiplet occurs only at the Fermi level, indicating that these states develop spontaneously due to electron-electron exchange interactions. Individual spectra as the (4)-fold degenerate multiplet crosses the Fermi level (Fig. 1e) demonstrate the change in relative amplitude of the two split LL peaks, which match well to the above filling factor assignments. We quantitatively extract the magnitude of the exchange gaps for each broken symmetry state, which reaches a maximum of $\Delta_{exch} = 650$ µeV and is similar for all integer $\tilde{v}$ [Fig. 1g (blue)]. Exchange interactions also enhance the gap $\Delta_{str}$ between LLs that are already split by strain by a similar amount, as shown in Fig. 1g (red). Thus, we observe interaction effects at all integer filling factors, including for states at odd filling factors that are not split by strain and have not been previously reported.

To address the valley occupation at odd integer filling factors, we first perform large-scale imaging of the Landau orbits at the energies of two singly degenerate $N = 3$ LLs (Fig. 2a-b,d). We observe the same orientation of wave function anisotropy for both of the singly degenerate LL peaks, which confirms that they arise from the subspace composed of the two valleys at opposite momenta. These LL wave functions, which show up as multiple sets of concentric ellipses, are centered around surface (circles) and sub-surface defects (arrows), as labeled in the concurrently measured topography (Fig. 2c). To first order, both types of defects can be treated as short-range potentials, which shift the energy of the single cyclotron orbit within each LL that has weight at the defect site[15]. Therefore, measurements performed at the unperturbed LL energies show decreased conductance in the shape of a single electronic state, allowing for the direct imaging of isolated Landau orbit wave functions.

High-resolution imaging around impurities allows us to use the two types of defects to independently demonstrate that the ground state at $\tilde{\nu} = 1$ is valley-polarized. Although the potential induced at the surface by sub-surface and surface impurities are comparable in strength, they differ in spatial extent (see Supplementary Information) and therefore have strikingly different effects on the measured conductance maps (Fig. 2e-f). The atomic scale surface defect potential enables large momentum transfers that allow coupling between valleys, regardless of the nature of the ground state. In contrast, the smoother potential of a defect below the surface is sharp only relative to the magnetic length and does not induce valley mixing (see Supplementary Information). Therefore, periodic density modulations should be visible around a sub-surface defect if and only if the ground state is a coherent superposition of valleys. The absence of such interference fringes in conductance maps around a sub-surface defect (Fig. 2e and zoom-ins in Fig. S7) reflects the lack of inherent charge modulation in the wave function and thus provides the first indication of a valley-polarized ground state.

Wave function maps in the vicinity of surface defects exhibit characteristically different behavior from the sub-surface defect maps, and do show interference patterns, but only in the vicinity of the defect (Fig. 2f). We explore the origin and detailed structure of these features below, which arise due to defect-induced valley mixing, and are not an indication of a valley-coherent ground state with intrinsic charge modulations. The spectra in Fig. 3a show exchange splitting of a (2)-fold degenerate LL and illustrate the shift in the LL peaks at the site of an isolated surface defect. Conductance maps with very fine spatial resolution taken at the energies of the two unperturbed, singly degenerate LL peaks are shown in Fig. 3b-c. Distinct vertical fringes with a wavevector corresponding to the separation between the two pockets at opposite momenta are stronger in Fig. 3b, whereas the nodes of the wave function are more visible in Fig. 3c. The differences between these two maps reflect the details of the energies of the states involved in the valley mixing caused by the surface defect (see Supplementary Information). We further confirm that these patterns result from disorder induced valley mixing by imaging similar interference even for a two-fold degenerate multiplet, when both valleys are completely occupied (see Supplementary Information).

We have developed a comprehensive theoretical model to better understand the interference between LL wave functions from different valleys, which confirms that the ground state at $\tilde{\nu} = 1$ is a valley-polarized ferroelectric. We construct an effective two-band Hamiltonian that captures the teardrop shape of the hole pockets of the Bi(111) surface states, and is in agreement with ARPES measurements. We then calculate the intrinsic wave functions for both possible ground states in the case of a singly degenerate LL and use them to compute the local density of states in the vicinity of both types of defects (see Supplementary Materials). This theory accounts for the nontrivial momentum-space spin texture of the hole pockets, which ensures valleys at opposite momenta have an overlapping spin component that permits mixing by a spin-independent disorder potential (Fig. 1b-c). A schematic of the energies used in our calculations for the $N = 3$ LL is shown in Fig. 3d, where the degeneracies of the different levels correspond to the spectra in Fig. 3a. The guiding centers that count the usual orbital degeneracy within a LL are indexed by $m$; the shifted zero angular momentum ($m = N$) states that have weight at the defect site are also included in the schematic above. The surface defect potential couples states with an energy difference comparable to the potential strength; the specific states from all six valleys that can interact within the framework of this model are marked by the dashed box in Fig. 3d. Numerical simulations of the local density of states for a valley-polarized ground state that incorporate this valley mixing match well to the experimental data (Fig. 3e-f), capturing the differences between each of the singly degenerate LLs. The strongest mixing occurs between the states coupled by arrows in Fig. 3d, which gives rise to the prominent vertical interference fringes. In contrast, simulations assuming a valley-coherent state (Fig. 3g-h) display periodic charge modulations that extend well beyond the vicinity of the defect and are inconsistent with our observations. Thus, in the case of a surface defect, theoretical simulations corroborate the valley-polarized nature of a singly degenerate LL. Moreover, this ground state is independently confirmed by comparing our experimental data to calculations around a sub-surface defect (see Supplementary Information).

Physically, the preference for valley polarization can be understood to follow from the fact that intra-valley exchange is always stronger[6] than inter-valley exchange, consistent with our numerical calculations. However, the in-plane electric dipole that necessarily results from the lack of two-fold symmetry for an individual pocket in this valley-polarized wave function is challenging to observe in the current experiment. Direct visualization of the dipole moment is hindered by its small magnitude (~0.4 nm), which is only 0.2% of the wave function spatial extent (see Supplementary Information). We expect that spontaneous valley polarization leads to ferroelectric domains in the sample, but the inability to directly measure the dipole through wave function imaging precludes identification of the local order parameter. Although our spectroscopic measurements are capable of identifying ferroelectric domain walls, no such boundaries were detected in this experiment (see Supplementary Information).

The distinctive spatial patterns that we observe represent a different regime of wave function interference than is typically measured by STM. Whereas traditional quasiparticle interference[29] (QPI) is the result of elastic scattering of the Bloch states of a crystal due to a defect potential, the interference presented here involves scattering between individual cyclotron orbits in a magnetic field, where the surface defect can perturbatively couple states at different energies. The novel wave function mixing and the involvement of different valleys is evident from the conductance maps and the fast Fourier transforms (FFTs) of the fine interference patterns around surface defects (Fig. 4). The real space maps at the energies of the shifted counterparts of the singly degenerate LLs (Fig. 4a-b) are not a simple contrast reversal of the maps taken at the unperturbed LL peaks, but display additional diagonal interference patterns, which match well to the theoretical simulations in Fig. 4e-f. The corresponding FFTs (Fig. 4c-d,g-h) show multiple groups of peaks in the FFT that are arranged either in a line or in a diamond pattern, in contrast to isolated scattering wavevectors visible in typical QPI data. Each group is centered around wavevectors corresponding to the center-to-center distance between pairs of hole valleys, indicating that the defect couples states from all six valleys within a LL even though they have different energies.

The scattering patterns in the FFTs reveal several further details about the wave functions involved and their spin textures. The outer boundary of the groups in the FFTs (Fig. 4c-d,g-h) is due to the size, shape and relative angle between the valleys involved in the scattering process, while the number of high intensity points reflects the nodal structure of the wave functions and therefore depends on orbital index (see Supplementary Information). In addition, although the LLs corresponding to the B and C valleys occur at the same energy, the signal in the FFT for $Q_{A'B}$ is significantly weaker than for $Q_{A'C}$, because of the stronger overlap between valley spin textures in the latter case (see labels in Fig. 1a). We note that it matters only whether the pockets are adjacent, not the direction of their anisotropy; the signal around $Q_{B'A'}$ is strong, whereas that near $Q_{A'B}$ is weak. Thus, by comparing the intensity of different groups of peaks, we can qualitatively determine the relative spin overlap for states in different valleys.

Our experimental approach to image the fine features of Landau orbits, resulting in the identification of a new class of broken symmetry quantum Hall state, have broader applicability. The single valley-polarized quantum Hall state with emergent ferroelectricity studied here is also expected to form in other anisotropic 2D valley systems, such as the surface states of topological crystalline insulators[16, 30]. More broadly, our experimental approach to image Landau level wave functions with a STM can also be extended to identify other exotic correlated states, perhaps including phases containing skyrmions[31] or fractional quasiparticles[32] that can form in high magnetic fields.

**Methods**

Single Bi crystals were grown using the Bridgman method from 99.999% pure Bi that had been treated to remove oxygen impurities. The samples were cleaved in ultrahigh vacuum at room temperature, immediately inserted into a home-built dilution refrigerator STM and cooled to cryogenic temperatures. Measurements were performed at 250 mK using a W tip. Spectra and conductance maps were acquired using a lock-in amplifier with AC rms excitation $V_{rms}$ = 30 µV for Fig. 1 and Fig. 3a, and $V_{rms}$ = 74 µV for the remainder of the data in Figs. 2-4. The setpoint voltage was $V_{set}$ = -400 mV and the setpoint current was $I_{set}$ = 5 nA. The data that support the plots within this paper and other findings of this study are available from the corresponding author upon reasonable request

**Acknowledgements**

We would like to thank I. Sodemann and L. Fu for helpful discussions. Work at Princeton has been supported by Gordon and Betty Moore Foundation as part of EPiQS initiative (GBMF4530), DOE-BES grant DE-FG02-07ER46419, ARO-MURI program W911NF-12-1-046, NSF-MRSEC programs through the Princeton Center for Complex Materials DMR-142054, NSF-DMR-1608848, Eric and Wendy Schmidt Transformative Technology Fund at Princeton, and by a NSF Graduate Research Fellowship (M.T.R.) and a Dicke fellowship (B.E.F.). Work at Austin was supported by DOE grant DE-FG03-02ER45958 and by Welch Foundation grant TBF1473. The work of F.W. at Argonne is supported by Department of Energy, Office of Basic Energy Science, Materials Science and Engineering Division.

**Author Contributions**

M.T.R., B.E.F., H.D., A.G. and A.Y. designed and conducted the STM measurements and their analysis. F.W. and A.H.M performed the theoretical modeling and simulations. H.J. and R.J.C. synthesized the samples. All authors contributed to the writing of the manuscript.

**Competing Financial Interests**

The authors declare no competing financial interests.

**Figure captions:**

**Figure 1 | Exchange splitting and broken symmetry states at odd integer Landau level (LL) filling. a**, Fermi surface for the Bi(111) surface states showing the electron pockets (gray) and the six anisotropic hole pockets, color coded to match the schematics below. **b,c**, Teardrop-shaped hole valleys at opposite momenta. The nontrivial spin textures are marked by arrows

denoting the magnitude and direction of the in-plane spin component, and filled (empty) circles denoting the +(-)z out-of-plane component. **d**, Landau fan diagram indicating the dispersion of the electron-like (positive slope) and hole-like (negative slope) LLs as a function of magnetic field. The LLs are pinned to the Fermi level until they are completely occupied. The degeneracy of the hole LLs at select magnetic fields is labeled in parentheses; a full lifting of the hole valley degeneracy is seen for both a (4)-fold and (2)-fold degenerate LL due to exchange interactions at the Fermi level. **e**,**f**, Individual spectra at particular magnetic fields, marked by arrows in the LL fan diagram in **(d)**, which highlight the relative amplitudes of the various split LL peaks. The number of occupied hole LLs within a given orbital index is specified by $\tilde{\nu}$. **g**, Magnitude of the exchange gap, $\Delta_{exch}$ (blue) as the degeneracy of the (4)-fold multiplet is lifted around the Fermi level. Exchange enhancement of the strain induced splitting, $\Delta_{str}$, when the Fermi level lies between the (4)-fold and the (2)-fold degenerate peaks (red).

**Figure 2 | Imaging LL wave functions at odd integer filling factors. a**,**b**, Large-scale conductance maps at the energies of the two singly degenerate LLs, showing the anisotropy and distinct nodes of individual wave functions pinned to defects. **c**, Simultaneously measured topography of the Bi(111) surface, showing two atomic-scale surface defects (circled). Sub-surface defects are not directly visible in the topography, and their positions, marked with arrows, are inferred from the centers of Landau orbits in the conductance maps. **d**, Representative spectrum in this region of the sample, showing the exchange splitting of the (2)-fold degenerate multiplet into two singly degenerate LLs [blue and red dots correspond to energies of maps in **(a-b)**]. **e**,**f**, High-resolution conductance maps around a sub-surface and a

surface defect showing the wave function for the $N = 3$ LL. The maps in (**e-f**) have been rotated to align the vertical axis with the crystallographic [$\bar{1}$10] direction as labeled in (**c**).

**Figure 3 | Interference fringes around a surface defect due to mixing of valley-polarized states. a**, Spectra taken on a pristine Bi(111) surface (black) and on a surface defect (orange), demonstrating the shift in the LLs due to the defect potential. **b,c**, Conductance maps at the energies of the two singly degenerate LL peaks, showing distinct vertical interference fringes in the vicinity of the defect. **d**, Schematic of the $N = 3$ LL energies, taking into account the lifting of the six fold valley degeneracy due to strain and exchange interactions, as well as the energy shift of the $m = N$ state due to the defect. States in red and blue are singly degenerate, whereas those in yellow and green are (2)-fold degenerate. The pointlike defect introduces important valley mixing terms between states within the dashed box, with the most prominent interference processes marked by arrows. This diagram does not include any energy renormalization due to defect-induced mixing, but valley mixing is completely accounted for in all numerical calculations. **e,f**, Theoretical simulations of the local density of states (LDOS) at the unperturbed LL energies [3]-[4] for a valley-polarized ground state with defect-induced mixing, which shows good agreement with the data. **g,h**, In contrast, similar calculations of the LDOS assuming an intrinsic valley coherent ground state do not match our observations.

**Figure 4 | Diagonal interference patterns reflecting the nodal wave function character and spin texture of the pockets. a,b**, Conductance maps at the shifted energies for $\tilde{\nu} = 1$ display additional diagonal interference features at the same field and around the same defect as in Fig. 3. **c,d**, Fourier transforms of the experimental data in (**a-b**) demonstrating the coupling of all six

hole valleys. The centers of the boxes mark the wavevectors corresponding to the center-to-center distance between pairs of valleys (Fig. 1a). **e,f**, Corresponding numerical simulations of the LDOS assuming a valley-polarized ground state, which match exceptionally well to the data in (**a-b**). **g,h**, Fourier transforms of the theoretical simulations in (**e-f**). The intensity of the peaks within a group is stronger for smaller energy differences and larger spin overlaps between the involved valleys.

Figure 1

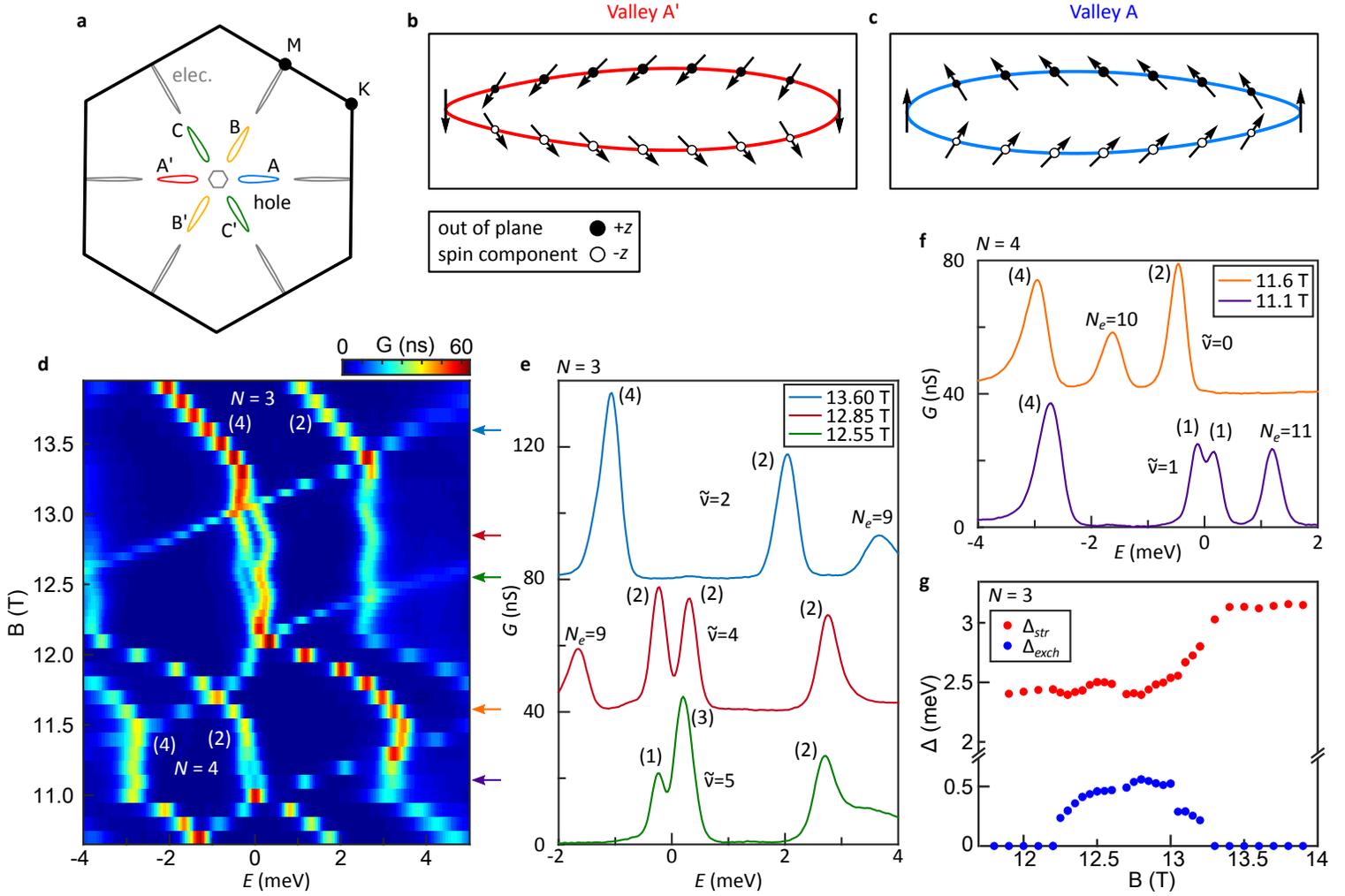

Figure 2

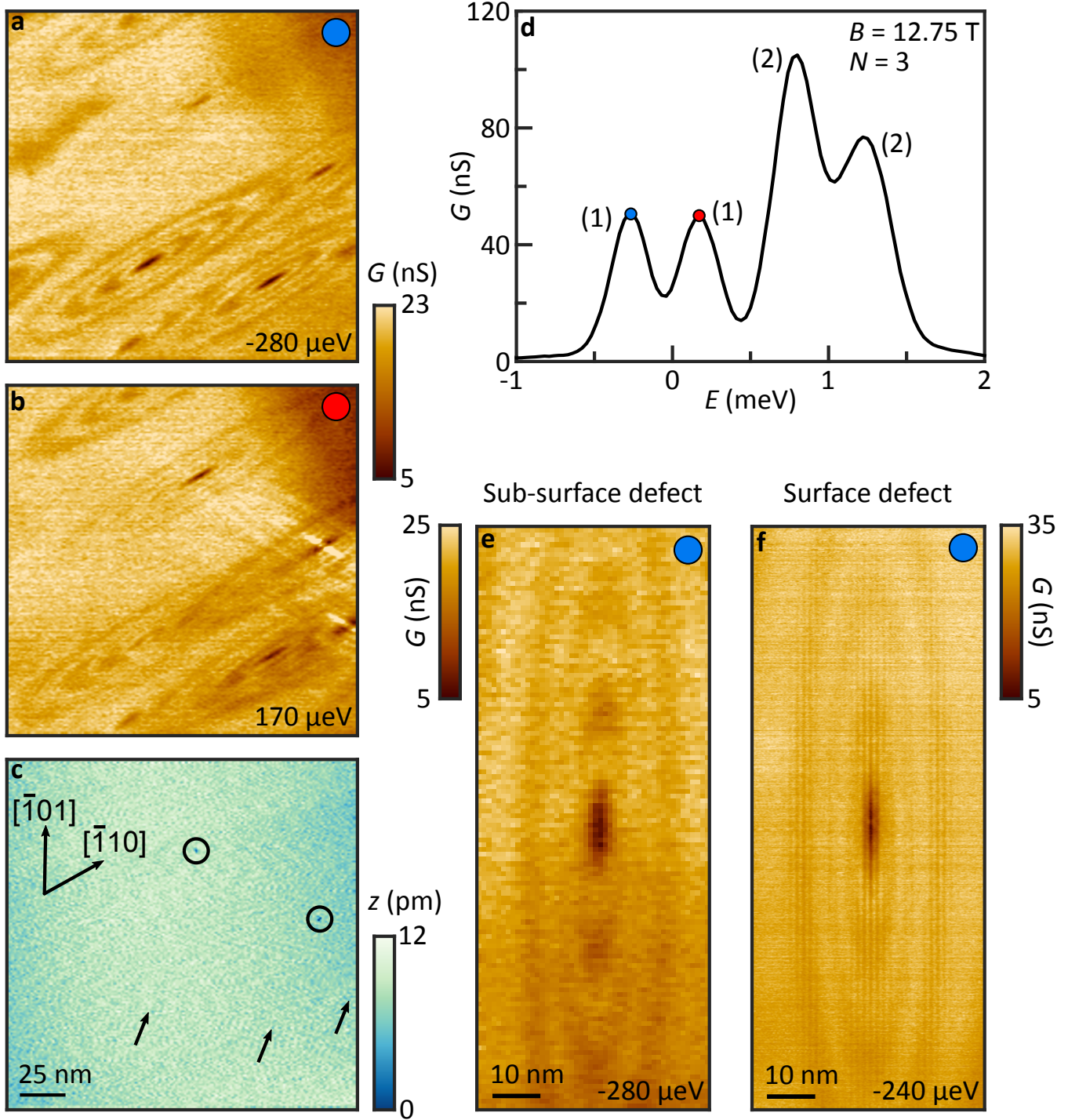

Figure 3

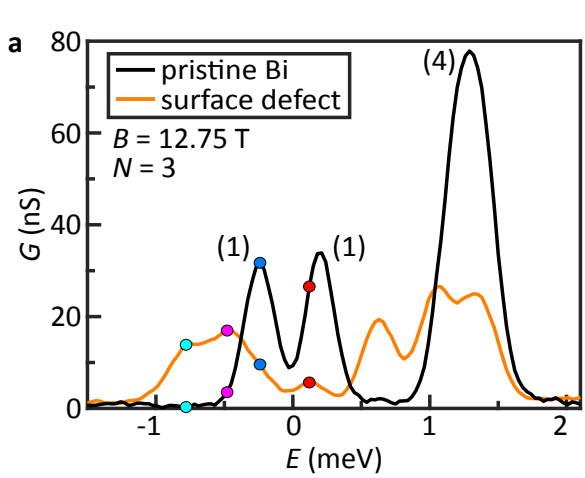
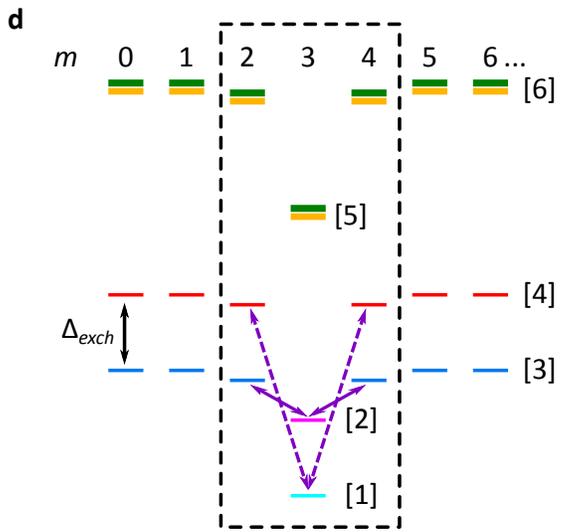
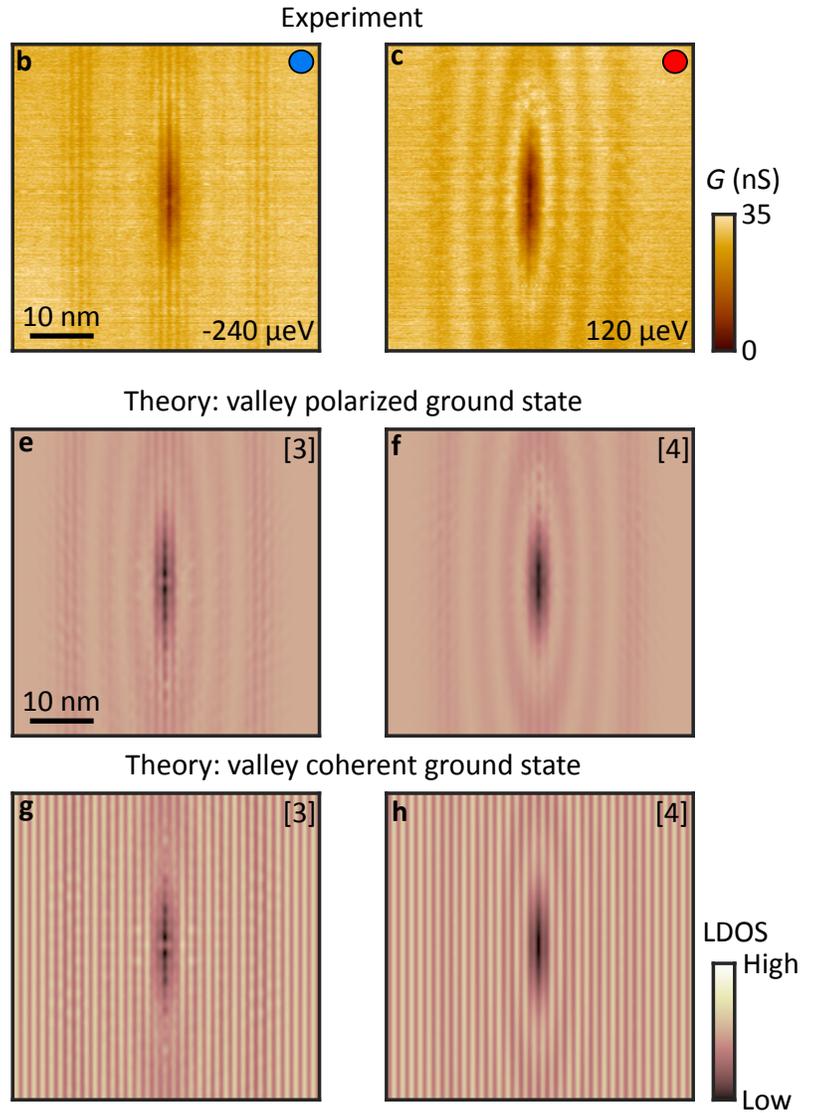

# Figure 4

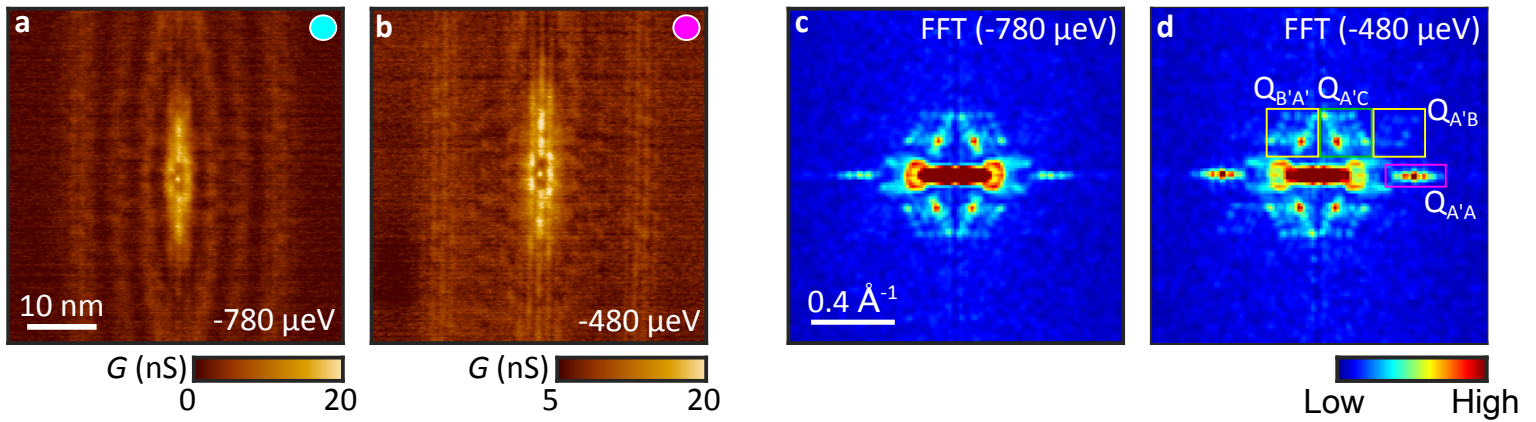

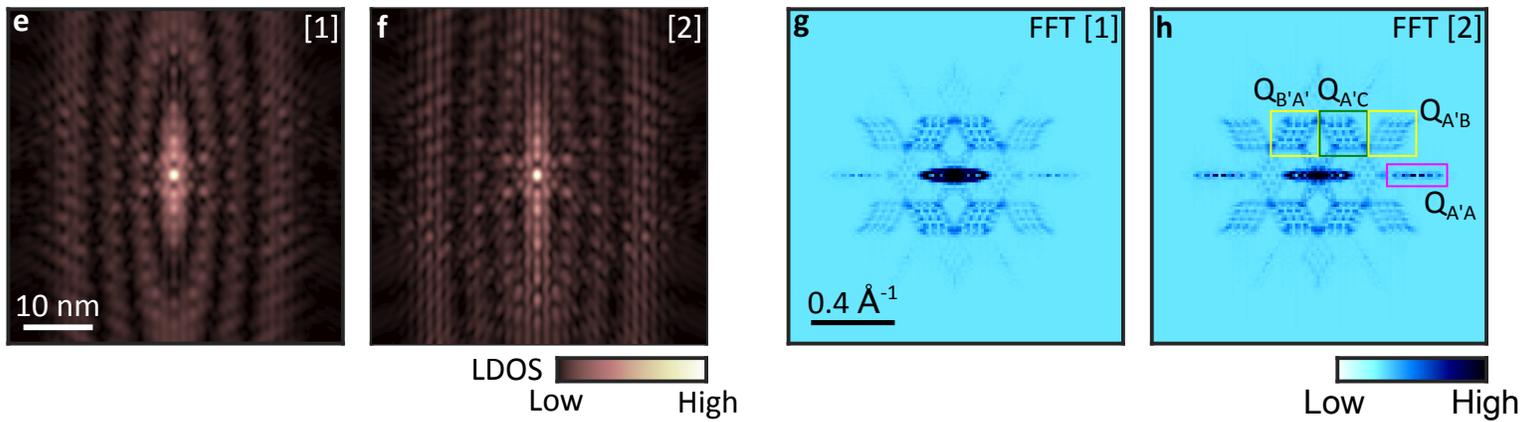